%% file: main.tex
\documentclass[10pt, a4paper, oneside, reqno]{amsart}

\input{style_P}
\input{command}

\usepackage{url}            

\usepackage{algorithm}
\usepackage{algorithmic}

 \usepackage[foot]{amsaddr}

\usepackage{subfig}

\usepackage{algorithmic}
{

{
		\theoremstyle{plain}
		
	}
	{
		\theoremstyle{plain}
		
	}
	{
	\theoremstyle{plain}
	
}
	{
		\theoremstyle{plain}
		
	}
	{
		\theoremstyle{plain}
		
	}
{
		\theoremstyle{plain}
		
	}
{
		\theoremstyle{plain}
		
	}}

\title[]{Safe multi-agent deep reinforcement learning for joint bidding and maintenance scheduling of generation units}

\author[1]{Pegah Rokhforoz$^1$}
\author[2]{Olga Fink$^{1}$}
\address[1]{Chair of Intelligent Maintenance Systems, ETH Zurich, Switzerland. Corresponding author, email address:ofink@ethz.ch.}

\begin{document}

\maketitle

\begin{abstract}
This paper proposes a safe reinforcement learning algorithm for generation bidding decisions and unit maintenance scheduling in a competitive electricity market environment. In this problem, each unit aims to find a bidding strategy that maximizes its revenue while concurrently retaining its reliability by scheduling preventive maintenance. The maintenance scheduling provides some safety constraints which should be satisfied at all times. Satisfying the critical safety and reliability constraints while the generation units have an incomplete information of each others' bidding strategy is a challenging problem. Bi-level optimization and reinforcement learning are state of the art approaches for solving this type of problems. However, neither bi-level optimization nor reinforcement learning can handle the challenges of incomplete information and critical safety constraints. To tackle these challenges, we propose the safe deep deterministic policy gradient reinforcement learning algorithm which is based on a combination of reinforcement learning and a predicted safety filter. The case study demonstrates that the proposed approach can achieve a higher profit compared to other state of the art methods while concurrently satisfying the system safety constraints. 
\end{abstract}

\section{Introduction}
Finding an optimal bidding strategy and maintenance scheduling of generation units in an electricity market would lead to a higher profit and an improved reliability of the entire system. The electricity market creates competition among the units since the profit of each unit depends on the market clearing price which is determined by the system operator (ISO), while the market clearing price is impacted by the bidding strategies of all units \cite{wang2017strategic}. This problem can be categorized as a multi-agent bi-level decision-making problem \cite{du2021approximating}. The units, which are referred to as agents, are strategic and price-takers. In the first level, the agents decide about their strategies individually. In the second level, the ISO clears the market and obtains the power generation amount for each unit such that the demand of the system can be satisfied in all periods of time \cite{kohansal2020strategic,zhang2010competitive}.

In this competition, the main aim of the unit is to maximize its profit by choosing the best bidding strategy. In addition to this goal, increasing the reliability and safety is another important objective that should be considered by the unit \cite{yildirim2016sensor}. In order to retain the reliability, the units need to perform preventive maintenance actions that take a defined period of time enabling, thereby, to prolong the lifetime of the units. However, performing maintenance imposes maintenance costs on the units. In this case, the units need to obtain their optimal maintenance scheduling which makes a trade-off between increasing reliability and decreasing the maintenance cost of the system \cite{chen1991optimal,volkanovski2008genetic}. Hence, in summary, the goal of the unit in the electricity market can be considered as finding the optimal bidding strategies and maintenance schedule which maximizes the system's profits as well as reliability and minimizes the maintenance cost.

Nash equilibrium (NE) is one typical solution to the competition problem where no unit can increase its profit by changing its strategies while the other units are not changing their strategies \cite{song2002nash}. Many papers address the optimal bidding of the electricity market using the game theory approach \cite{ye2019incorporating, dai2016finding,wang2017strategic,zhang2010restructured,pozo2011finding}. In these research studies, increasing reliability and maintenance schedules are, typically, not considered in the objective functions of the units. In fact, finding the NE of this problem is a challenging problem since the units are neither aware of the bidding of others nor of the market-clearing price of the system. In addition, another challenge is that the electricity demand should be satisfied at all time even when some units are out of operation and are performing maintenance. This imposes some constraints on the maintenance scheduling of units and the NE of the game should satisfy these critical constraints. Hence, the units face the incomplete information game associated with some uncertainties and constraints \cite{du2021approximating}. This is a challenging problem and needs some coordination among the units \cite{marwali1999long,pandzic2011yearly}

Besides the works that address the bidding strategies of units, many papers address the distributed maintenance scheduling problem \cite{feng2009competitive, min2013game}. These papers proposed optimization approaches in which the units decide about their maintenance scheduling. To handle the challenges of electricity demand satisfaction, an algorithm based on the coordination and negotiation between ISO and units is proposed in \cite{rokhforoz2021multi, conejo2005generation}. These papers assume that units send their marginal cost as the bidding price to the ISO. Therefore, obtaining the optimal bidding strategies is not their goal. 

All of the above works assume that the units have complete information of the market-clearing price as well as the bidding of their competitors. Thereby, they can solve the optimization problem and reach their NE. It is worth noting that this is a limiting assumption in the real market system.

Reinforcement learning (RL) is a promising tool to handle challenges of incomplete information and system uncertainty \cite{sutton2018reinforcement}. In the case of an electricity market, this incomplete information and uncertainty are due to the unknown information about the actions of other units as well as the market clearing price. 

Q-learning is one of the typical RL algorithms which applies look-up tables to find the optimal strategy in each state of the system\cite{watkins1992q}
. However, since a bidding strategy in  electricity markets is a continuous action, this method has a high computational time \cite{millan2002continuous}. Deep Q-learning can tackle this challenge by using neural networks as function approximators for obtaining the strategies of the units \cite{arulkumaran2017deep,xu2019deep}. Some research studies address the bidding strategies using deep RL algorithm  \cite{du2021approximating,liang2020agent,ye2019deep}. However, they do not consider maintenance scheduling in addition to the bidding strategies.

As mentioned above, applying maintenance scheduling as one of the aims of the generation units imposes some critical constraints on the system such as electricity demand satisfaction and allocating the time for performing preventive  maintenance. These constraints related to system safety and reliability should be satisfied at all time periods. Unfortunately, deep RL cannot ensure that all the system constraints can be satisfied during the training of the units.

One way to solve this issue is to combine a predicted safety filter with a RL algorithm \cite{wabersich2021predictive}. In the proposed approach, a safety filter is responsible for satisfying system safety constraints and finding  strategies which are as close as possible to the RL strategies while also satisfying all the system constraints. In other words, if RL strategies violate some constraints of the system, the predicted safety filter adjusts the decisions of the units to ensure that the constraints are satisfied. 

In this paper, we propose to define the joint bidding and maintenance scheduling of generation units in the electricity market system as a safe deep RL problem. The proposed approach is able to satisfy the safety constraints and also handle the challenges of incomplete information due to the unknown bidding strategies of other units. In this approach, the units do not need to know the model and the bidding strategies of other units. They can update their strategies by getting feedback and reward from the electricity market system in each time step of the learning algorithm. In other words, at each time step, the units obtain their strategies and send them to the ISO. Then, they obtain their profit based on the information that they get from ISO, and update their policy. The main contributions of this paper are as follows:

1) We propose to formulate the maintenance scheduling and electricity bidding process of multiple generation units as a  bi-level optimization problem. In the first level, the units aim to obtain the strategies which maximize their reward function while satisfying the system safety constraints related to the maintenance scheduling of agents. In the second level, the ISO clears the market and obtains the electricity price and power generation of each unit such that the total cost of the system is minimized and all the network constraints are satisfied.

2) We develop a multi-agent deep RL using the deterministic policy gradient (DDPG) algorithm to obtain the maintenance scheduling and bidding price such that the reward function of the agents is maximized without considering the safety constraints. This algorithm can handle the high-dimensional continuous action and state space. Moreover, it can tackle the non-stationary and uncertainties of the environment where each unit is not aware of the decisions of other units.

3) We design a predicted safety filter to guarantee that all the safety constraints are satisfied. This filter ensures that during the training as well as testing of the RL algorithm the units can perform maintenance for some specified time intervals. Moreover, this filter provides the conditions which ensure that the load of the system can be satisfied at all times by limiting the number of units that can perform maintenance simultaneously.

4) The proposed methodology can be applied online to the real system since the safety filter makes sure that even during the training of the learning algorithm all the safety and demand constraints can be satisfied. In other words, the agent does not need to learn its strategies offline and we do not need to build a virtual electrical market environment which is associated with many uncertainties. 

The rest of the paper is organized as follows. The preliminaries on RL methodology and algorithm are introduced in Section \ref{sec:preliminary}. The units' and ISO's objective functions are formulated in Section \ref{sec:problem formulation}. The solution method based on a combination of RL and predicted safety filter is proposed in Section \ref{sec:solution}. Simulation results of the case study are presented in Section \ref{sec:result}. Concluding remarks are made in Section \ref{sec:conclusion}.

\subsection{Indices and sets}
$\mathcal{N}$: set of units ${\{1,\cdots,N}\}$ indexed by i. 

$\mathcal{N}_{s}$: set of samples ${\{1,\cdots,N_{s}}\}$ indexed by j.

$\mathcal{T}$: set of operational intervals ${\{1,\cdots,T}\}$ indexed by t.

\subsection{Parameters}
$t$: Number of time period 

$\powermax$: maximum power generation of generation unit $i$ ($MW$)

$\powermin$: minimum power generation of generation unit $i$ ($MW$)

$\marginalcost$: marginal cost of generation unit $i$ ($\frac{\$}{MW}$)

$\biddingmax$: maximum bidding of generation unit $i$

$\rampup$, $\rampdown$: ramp up/down of generation unit $i$

$M$: Maximum number of generation units that can perform maintenance simultaneously

$H_{i}$: Required duration of preventive maintenance of generation unit $i$

$D_{i}$: Minimum duration of a single maintenance action of generation unit $i$

$c_{i}$: Maintenance cost of unit $i$ 

\subsection{Variables}

$g_{i}(t)$: power generation of generation unit $i$ at time $t$ ($MW$)

$\price$: electrical market price at time $t$ ($\frac{\$}{MW}$)

$\maintenancet$: maintenance decision of generation unit $i$ at time $t$

$\biddingt$: bidding decision of generation unit $i$


\section{Preliminary on RL algorithm}
\label{sec:preliminary}

The aim of RL is to maximize the total discounted reward function where the environment can be modeled as a Markov decision process (MDP) \cite{sutton2018reinforcement}. An action-value function is defined as follows:

\begin{equation}
    Q_{\pi}(s({t}),a({t}))= {\mathbb E}_{\pi}\Big[\sum\limits_{k=0}^{N_{T}}\gamma^{k}r({t+k+1})|s({t}),a({t}),\pi\Big],
\end{equation}
where $s({t})\in\mathcal{S}$ and $a({t})\in\mathcal{A}$ are the state and the action at time $t$. The policy is the probability of choosing action $a$ given state $s$. The reward and discount factor are expressed by $r$ and $\gamma$ where $\gamma\in[0,1)$. The expected value of the reward over policy $\pi$ is given by ${\mathbb E}_{\pi}[\cdot]$. The aim of the RL algorithm is to find the policy which maximizes the action-value function as follows:

\begin{equation}
    Q^{*}(s_{t},a_{t})= \max\limits_{\pi}Q_{\pi}(s_{t},a_{t}).
    \label{eq:q learning}
\end{equation}
The conventional RL methods such as Q-learning can solve the action-value function \eqref{eq:q learning} using a look-up table. In this case, a table is established to store the action-value function values for all possible state-action pairs. One typical way to update a table is to implement the temporal difference method \cite{menache2005basis} in which the value-action function is updated as follows:
\begin{equation}
    Q_{\pi}^{l+1}\big(s({t}),a({t})\big)=r({t})+\gamma\max\limits_{a(t+1)}Q_{\pi}^{l}\big(s({t+1}),a({t+1})\big),
\end{equation}
where $l$ is the iteration index. 

However, in cases where either the state or action space is continuous, Q-learning faces the ``curse of dimensionality'' problem. Deep RL is one way to tackle this challenge. In this method, instead of a look-up table, a neural network is trained to estimate the action-value function. This network can create a continuous mapping from the state and action pairs to the action-value function.

\section{Problem formulation}
\label{sec:problem formulation}
In this section, we formulate the optimization problem of generation units and ISO for the electrical market with a uniform price. In the proposed model, the generation units decide about their maintenance scheduling $u_{i}$ and bidding strategy $k_{i}$ while the ISO obtains the amount of power generation for each unit $g_{i}$ and the electrical market price $\lambda_{i}$ through the following optimization problems. 

\subsection{Maintenance model of generation units }
\label{sec:maintenance}
In this part, inspired by \cite{conejo2005generation}, we model the unit maintenance cost and its corresponding constraints. In the proposed model we assume that unit $i$ is charged the maintenance cost $c_{i}$ at each time while performing maintenance and seeks to find the strategy that minimizes the total maintenance cost. In addition, we assume that during the decision horizon $\mathcal{T}$ unit $i$ needs to perform maintenance at least for the total duration of $H_{i}$. This preventive maintenance activity ensures that the generation unit can retain the system reliability. We assume that $H_{i}$ is determined by the safety regulation for each unit which is set based on the generation unit's data and model. Based on these preventive maintenance activities, we impose the following constraint:

\begin{equation}
    \SUM\maintenancet\geq{H_{i}},\quad \N.
    \label{eq:maintenance constraint1}
\end{equation}

In addition, there is another constraint which represents the condition that if the maintenance of each unit starts at time $t$, it should be completed, which takes $D_{i}$ periods of time. We formulate this constraint as follows:

\begin{equation}
\maintenancet-\maintenancetp\leq{\maintenanced},\quad\TA,\N.
\label{eq:maintenance constraint2}
\end{equation}
It is worth to mention that these two constraints, \eqref{eq:maintenance constraint1} and \eqref{eq:maintenance constraint2}, imply that for the whole decision horizon of time $t\in\mathcal{T}$, unit $i$ can perform maintenance several times with an interruption between them, provided that each maintenance action takes $D_{i}$ periods of time and the total period of maintenance is larger than the required duration $H_{i}$ of preventive maintenance of generation unit $i$.

In the following, we propose the optimization model of generation units using the above discussion.

\subsection{Optimization model of generation units}
In this section, we propose the optimization problem in which the goal of the units is to obtain their optimal bidding and maintenance scheduling strategies which maximize their revenue and also satisfy the maintenance constraints (Equations \eqref{eq:maintenance constraint1} and \eqref{eq:maintenance constraint2}). In this optimization problem formulation for clarity reasons and without loss of generality, we assume that each producer has only one generation unit. In addition, we assume that the start up and shut down costs are zero and each unit has a linear marginal cost. The optimization problem is formulated as follows:

\begin{equation}
\begin{aligned}
    \max\limits_{\bidding, \maintenance}\SUM&\Big(\price\power-\marginalcost\power-c_{i}\maintenancet\Big)\\
    \text{subject to}:&\\
    & {\text{A}_{1}}:{1}\leq\bidding(t)\leq{\biddingmax},\quad\TA,\\
    &{\text{A}_{2}}:\SUM\maintenancet\geq{H_{i}},\\
     &{\text{A}_{3}}:\maintenancet-\maintenancetp\leq{\maintenanced},\quad\TA,\\
     &{\text{A}_{4}}:\maintenancet={\{0,1}\},\quad\TA,
    \end{aligned}
    \label{eq:agent model}
\end{equation}
where the first term in the objective function $\price{g_{i}(t)}$ expresses the revenue that unit $i$ obtains by producing and selling the power. The second term $\marginalcost{g_{i}(t)}$ and the third term $c_{i}\maintenancet$ are the marginal and the maintenance costs of unit $i$. Constraint $\text{A}_{1}$ expresses the limit of the strategic bidding variable. In fact, each unit chooses a bidding strategy $\biddingt$ and offers $\biddingt\marginalcost$ where $\biddingt=1$ means that unit $i$ offers its actual marginal cost and behaves competitively, and ${1}<{\biddingt}\leq{\biddingmax}$ means that unit $i$ behaves strategically. Constraints $\text{A}_{2}$ and $\text{A}_{3}$ are the maintenance constraints which have been explained in Section \ref{sec:maintenance}. Constraint $\text{A}_{4}$ denotes that the maintenance decision is a binary variable where $u_i(t)=1$ indicates that the unit performs maintenance at time $t$.

In problem \eqref{eq:agent model}, the price of the market $\price$ and the power generation of units $g_{i}(t)$ are obtained through the ISO optimization problem which is described in the following section.

\subsection{ISO optimization model}
In this section, we derive the optimization formulation of ISO where the maintenance decision $\maintenance$ and bidding strategy $\bidding$ of the units are obtained through Eq.\eqref{eq:agent model}. ISO optimization aims to minimize the cost of the market while satisfying the load demand of the system. To achieve this aim, we formulate the problem as follows:

\begin{equation}
\begin{aligned}
    \min\limits_{g}\SUM\SUMI&\biddingt\marginalcost{g_{i}(t)}\\
    \text{subject to}:&\\
    &{\text{C}_{1}}:\SUMI\power=d(t):\quad\price,\quad\TA,\\
    &{\text{C}_{2}}:{(1-\maintenancet)\powermin}\leq{g_{i}(t)}\leq{(1-\maintenancet)\powermax},\quad\N,\TA,\\
    &{\text{C}_{3}}:\power-\powerp\leq{\rampup},\quad\N,\TA,\\
    &{\text{C}_{4}}:\powerp-\power\leq{\rampdown},\quad\N,\TA,
    \end{aligned}
        \label{eq:ISO}
\end{equation}
where the objective function is to minimize the total costs of the units. The clearing price of the system $\lambda$ is the dual multiplier of Constraint $\text{C}_{1}$. Constraint $\text{C}_{1}$ expresses the demand-supply balance constraint. Constraint $\text{C}_{2}$ denotes the maximum and minimum power that unit $i$ can produce. In fact, this constraint implies that when the unit is performing maintenance, its production output is zero. Constraints $\text{C}_{3}$ and $\text{C}_{4}$ express the ramp up and ramp down constraints of unit $i$.

In the following section, we explain our proposed methodology to solve problems \eqref{eq:agent model} and \eqref{eq:ISO}.

\section{Solution methodology: safe multi-agent deep RL}
\label{sec:solution}

\begin{figure}[!t]
\centerline{\includegraphics[width=\columnwidth]{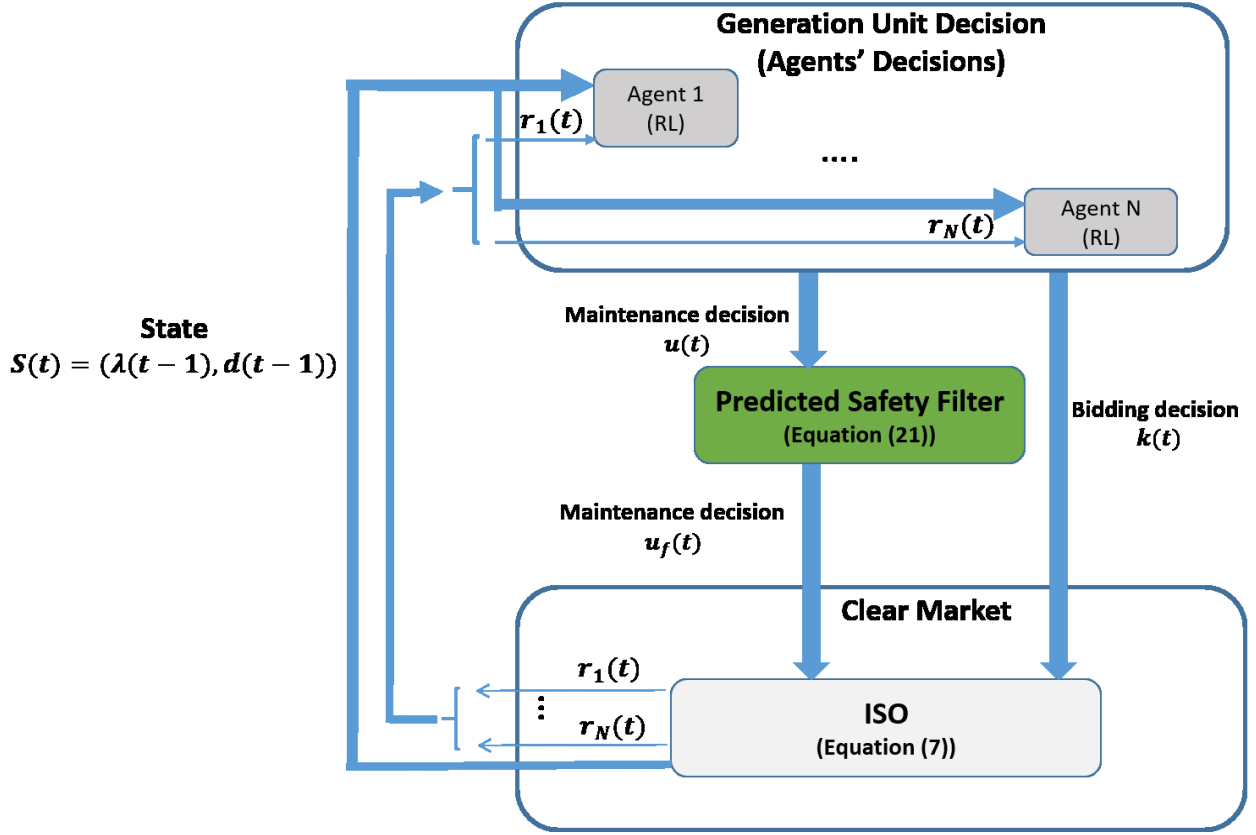}}
\caption{Schematic of the proposed solution methodology.}
\label{fig:schematic}
\end{figure}

As we can see from the problem formulation in Section \ref{sec:problem formulation}, we face a bi-level optimization problem. In this problem, in the upper level, the units aim to maximize their reward by obtaining their optimal bidding strategies and maintenance schedule. Also, the objective function of problem \eqref{eq:agent model} depends on the price $\price$ of the market which can be obtained through \eqref{eq:ISO}. Since the optimization of \eqref{eq:ISO} depends on the decision variables of all units, $\lambda(t)$ is the function of all other units' decisions. The dependency of $\lambda(t)$ on all the units' decision variables makes problem \eqref{eq:agent model} a challenging problem. On the one hand, each unit cannot obtain its decision variables without considering the decisions of other units. On the other hand, each unit is not aware of the decisions of other units. Hence, unit $i$ faces some uncertainty when solving problem \eqref{eq:agent model}. To solve these challenges, we propose a algorithm which is a combination of RL and a safety filter. The proposed algorithm comprises two parts. In the first part, the units learn the strategies of other units using RL algorithm. In the second part, all the critical safety constraints ($\text{A}_{2}$, $\text{A}_{3}$) are satisfied using the safety filter. In other words, the RL algorithms is able to find the strategies of the units without considering the safety constraints. Then, the units send their maintenance decisions to the safety filter and the filter obtains the decisions which are close to the decisions of the units while satisfying the safety constraints as well. In the next step, the bidding and maintenance decisions are sent to the ISO and it obtains the price and power generation of the units by clearing the market. Based on this information, the units update their strategies in order to gain more profit. The schematic of this approach is shown in Figure \ref{fig:schematic}. 

The formulation of the RL algorithm and the predicted safety filter are explained in the following sections.

\subsection{Deep RL using DDPG algorithm}
As mentioned above, the RL algorithm aims to obtain the strategies of the agents which consist of maintenance scheduling and bidding algorithm without considering the safety constraints $\text{A}_{2}$-$\text{A}_{3}$ of \eqref{eq:agent model}. Since the bidding strategy is a continuous action, we apply deep RL using 
deep deterministic policy gradient (DDPG) algorithm in which the two neural networks are used to obtain on the one hand the value function and on the other hand the policy for each of the units. To implement this algorithm, we need to define the state, the action, and the reward of each agent. Let us define the state of the system at time $t$ as follows:

\begin{equation}
    s(t)=(\lambda(t-1),d(t-1)),
    \label{eq:state}
\end{equation}
which  consists of the electrical price and the demand of the system at the previous time step. This state can be provided by ISO and describes the system environment. The action of unit $i$ can be defined as follows:

\begin{equation}
    a_{i}(t)=(\maintenancet, \bidding(t)),
        \label{eq:action}
\end{equation}
where $\maintenancet={\{0,1}\}$ and ${1}\leq\bidding(t)\leq{\biddingmax}$ are the maintenance and bidding strategies of generating unit $i$.

Let us consider the objective function of \eqref{eq:agent model}, the reward function of unit $i$ at time $t$ can be formulated as:

\begin{equation}
    r_{i}(t)=\price\power-\marginalcost\power-c_{i}\maintenancet.
        \label{eq:reward}
\end{equation}
Equation \eqref{eq:reward} implies that the reward of unit $i$ is the price that it obtains by selling its power to the market subtracting its generation and maintenance cost.

In the following, we explain the details of the DDPG algorithm and how the neural network weights are updated. As mentioned before, this algorithm comprises two neural networks which are estimating the action-value function and the policy of the units. They are also known as actor-critic networks. It should be mentioned that since we have $N$ units and each unit decides about its own strategy individually, we have an individual actor-critic network for each unit. In other words, we have $N$ actor-critic networks in which each unit updates the weights of the corresponding network using the DDPG algorithm. In this algorithm,  each agent first obtains its action $a_{i}(t)$ using the actor-network. In the next step, it executes the action and gets the reward $r_{i}(t)$ and the new state $s(t+1)$. Then, it stores the tuple $(s(t),s(t+1),a_{i}(t),r_{i}(t))$ and uses a mini-batch $(s^{j}(t),s^{j}(t+1),a_{i}^{j}(t),r_{i}^{j}(t))$ to update the actor-critic network weights. In the following, we explain the details of the DDPG algorithm.

\textbf{Critic network:} In the DDPG algorithm, for each unit, two critic networks are considered to estimate the action-value functions which are called target and behaviour critic networks. The inputs of the networks are the state $s(t)$ and the action of the units $a_{i}(t)$ which are defined as \eqref{eq:state} and \eqref{eq:action}, respectively. The output is the associated action-value function $Q_{i}(s(t),a_{i}(t))$. Let us consider $\theta'_{i}$ and $\theta_{i}$ as the weights of the target and the behaviour critic networks of unit $i$, respectively. The loss function for updating the weights of the behaviour critic networks is defined as follows:

\begin{equation}
L_{i}(\theta_{i})=\frac{1}{N_{s}}\sum\limits_{j\in\mathcal{N}_{s}}\big(Q_{i,target}^{j}(t)-Q_{i}(s^{j}(t),a_{i}^{j}(t);\theta_{i})\big)^{2},\quad\I,
\end{equation}
where the target network $Q_{i,target}$ is calculated as follows:

\begin{equation}
Q_{i,target}^{j}(t)=r_{i}^{j}(t)+\gamma\max\limits_{a_{i}^{j}(t+1)}Q_{i}(s^{j}(t+1),a_{i}^{j}(t+1),\theta_{i}'),\quad\J,
\label{eq:q_target}
\end{equation}
which in fact is the summation of the current reward $r_{i}(t)$ and the discounted value of the maximum action-value at the next time step $t+1$ which is obtained by the target network.

During training and updating of the weights of the networks, the target network is updated at a slower speed than the behaviour critic network. This condition helps to stabilize the learning process. The weights of the behaviour network for unit $i$ are updated as follows:

\begin{equation}
\theta_{i}\leftarrow\theta_{i}-\eta_{i}\nabla_{\theta_{i}}L_{i}(\theta_{i}),
\label{eq:critic weight}
\end{equation}
where $\eta_{i}$ is the learning rate of the behavior network. 

The weights of the target critic network for unit $i$ are updated as follows:

\begin{equation}
    \theta_{i}'\leftarrow(1-\tau_{i})\theta_{i}+\tau_{i}\theta_{i}',
    \label{eq:target critic network}
\end{equation}
where $\tau_{i}$ is the learning rate of the target network of unit $i$. 

\textbf{Actor network:} The actor network of unit $i$ is designed to obtain the optimal policy by considering the estimated action-value function, which is defined as: 
\begin{equation}
    \pi_{i}(s(t))=\arg\max_{a_{i}(t)}{Q_{i}(s(t),a_{i}(t))}.
\end{equation}

The input of the actor network is the current state $s(t)$ and the output is the action $a_{i}(t)$
which maximizes ${Q_{i}(s(t),a_{i}(t))}$. Let us define the loss function of the actor network as follows:

\begin{equation}
    L_{i}(\theta_{i}^{\mu})=-\frac{1}{N_{s}}\sum\limits_{\J}\big(Q_{i}(s^{j}(t),a_{i}^{j}(t);\theta_{i})\big)|_{a_{i}^{j}(t)=\mu_{i}(s^{j}(t),\theta_{i}^{\mu})}\big),
    \label{eq:loss actor}
\end{equation}
where $\theta_{i}^{\mu}$ is the weight of actor network and $\mu_{i}(s^{j}(t),\theta_{i}^{\mu})$ is the policy generated by the network. In fact, the goal of unit $i$ is to find the policy which maximizes ${Q_{i}(s(t),a_{i}(t))}$. Hence, by minimizing the loss function as \eqref{eq:loss actor} we can obtain the policy that maximizes ${Q_{i}(s(t),a_{i}(t))}$.

The weights of the actor network are updated as follows:

\begin{equation}
    \theta_{i}^\mu\leftarrow \theta_{i}^\mu-\eta_{i}^\mu\nabla_{\theta_{i}^{\mu}}L(\theta_{i}^{\mu}),
    \label{eq:actor network1}
\end{equation}
where $\eta_{i}^\mu$ is the learning rate of the actor network of unit $i$.

\begin{figure}[!t]
\centerline{\includegraphics[width=\columnwidth]{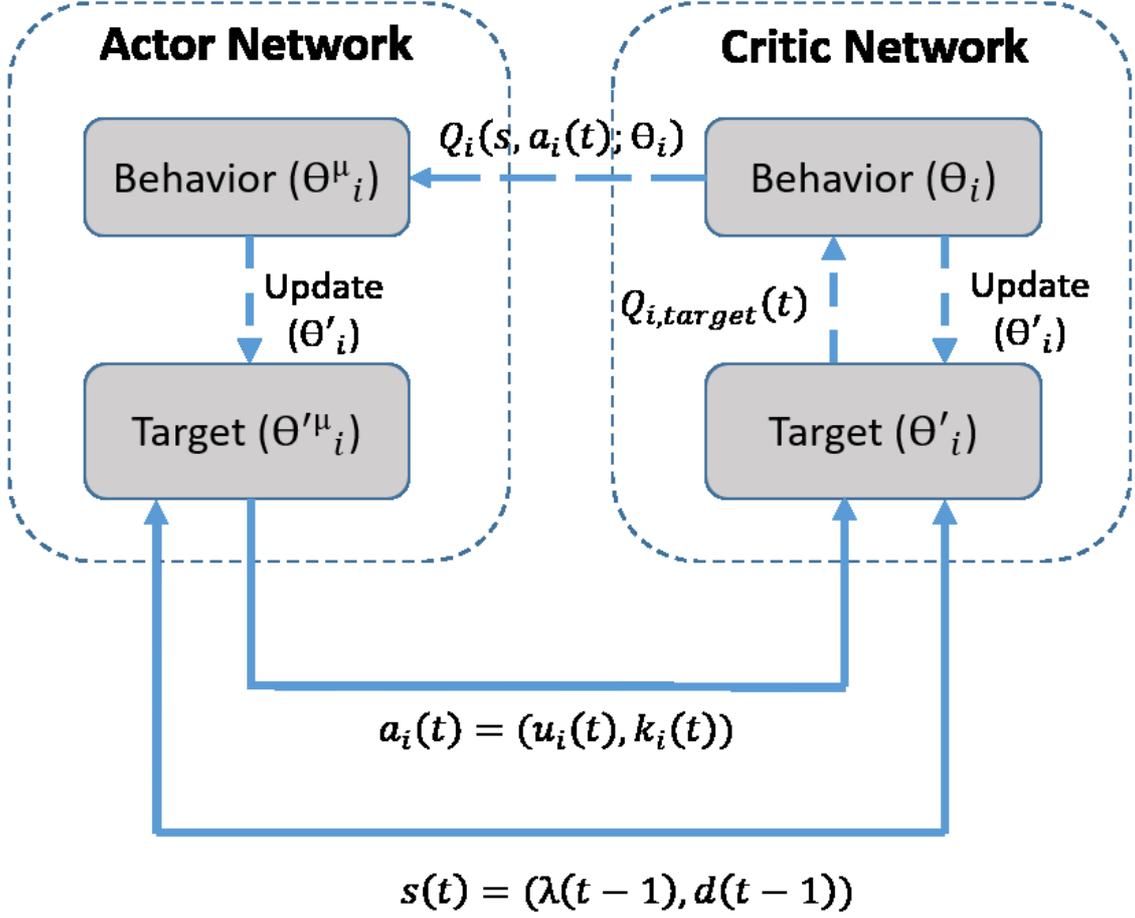}}
\caption{Schematic of the DDPG algorithm for unit $i$. Dash lines show the information that is used for updating the weights of networks.}
\label{fig:actor_critic}
\end{figure}

In the same way as for critic network, for the sake of stability, there is also a target actor network whose weights are updated as:
\begin{equation}
    \theta_{i}^{'\mu}\leftarrow(1-\tau^{\mu}_{i})\theta_{i}+\tau^{\mu}_{i}\theta_{i}^\mu,
    \label{eq:actor target network}
\end{equation}
The schematic of the DDPG algorithm for unit $i$ is depicted in Figure \ref{fig:actor_critic}.

\subsection{Predicted safety filter model}
As mentioned above, we aim to formulate the safety filter which obtains a maintenance decision close to the units' original decision while making sure that the updated decision guarantees that all the safety constraints of problem \eqref{eq:agent model} are satisfied for the predicted horizon time as well. To achieve this aim, at each instant $t$, the units send their decisions $\maintenancet$ to the predicted safety filter. Then, the predicted safety filter obtains maintenance decisions $u_{f,i}(t)$ which satisfy constraints $\text{A}_{2}$, $\text{A}_{3}$ for all periods of $\TS$. Moreover, one additional aim is to ensure that while several units perform maintenance simultaneously and are not able to produce any power in that period of time, the load at the network level can still be satisfied. To achieve this goal, we restrict the number of units that can perform maintenance simultaneously. By considering all of these restrictions, we propose to formulate the safety filter optimization as follows:

\begin{equation}
\begin{aligned}
      \min\limits_{\actionfilter}&\SUMI{\lVert{\actionfiltert-\maintenancet}\rVert}^{2}\\
      & {\text{S}_{1}}:\statenext=(1-\actionfilter(t'))\state+(1-\actionfilter(t')),\quad\TS,\N\\
      & {\text{S}_{2}}:\statefinal\geq{H_{i}},\N\\
      & {\text{S}_{3}}:\SUMI \actionfilter(t')\leq{M},\quad\TS,\\
     &{\text{S}_{4}}:\actionfilter(t')-\actionfilter(t'-1)\leq{\actionfilter(t'+D_{i}-1)},\quad\TS,\N\\
     &{\text{S}_{5}}:\actionfilter(t')={\{0,1}\},\quad\TS,\N
     \label{eq:safety}
      \end{aligned}
\end{equation}
where $x_{f,i}(0)=0$. The objective function implies that if the maintenance decisions of the units $\maintenancet$ can satisfy all the constraints, the output of the safety filter is the same as the units' decisions. Constraints $\text{S}_{1}$ and $\text{S}_{2}$ are equivalent to constraint $\text{A}_{2}$ of problem \eqref{eq:agent model}. In fact, $\state$ measures the duration that the unit performs maintenance which at the end of the decision horizon $T$ should be longer than $H_{i}$. Constraint $\text{S}_{3}$ ensures that maximum $M$ units can perform maintenance simultaneously, in which choosing the appropriate value of $M$ guarantees that at each time instant there is a sufficient number of units in the power grid system that can satisfy the electrical demand of the system. Constraints $\text{S}_{4}$ and $\text{S}_{5}$ are the same as $\text{A}_{3}$ and $\text{A}_{4}$ which imply that the units need $D_{i}$ time for performing a maintenance action and the decision of the predicted safety filter is a binary variable. 

Due to the nonlinear constraint $\text{S}_{1}$, Problem \eqref{eq:safety} is a nonlinear programming problem. Using big-M method, we convert \eqref{eq:safety} to a mixed integer linear programming (MILP) problem. Let us consider $Z_{i}(t')=\actionfilter(t')\state$ and add the following constraints to \eqref{eq:safety} as an equivalent of the nonlinear term:

    \begin{equation}
    \begin{aligned}
        &{0}\leq Z_{i}(t')\leq\actionfilter(t')M\\
        &Z_{i}(t')\geq{\state-(1-\actionfilter(t'))M}\\
        &Z_{i}(t')\leq{\state+(1-\actionfilter(t'))M},
        \label{eq:big m}
        \end{aligned}
    \end{equation}
        where $M$ is a large positive constant \cite{Fortuny_1981}.
        
Then, using \eqref{eq:big m}, we have the following MILP optimization:

\begin{equation}
\begin{aligned}
      \min\limits_{\actionfilter,Z_{i}}&\SUMI{\lVert{\actionfiltert-\maintenancet}\rVert}^{2}\\
      & {\text{S}_{1}}:\statenext=\state-Z_{i}(t')+(1-\actionfilter(t')),\quad\TS,\\
      & {\text{S}_{2}-\text{S}_{5}},\eqref{eq:big m}.\\
     \label{eq:safety1}
      \end{aligned}
\end{equation}

\subsection{Learning algorithm for electricity market bidding and maintenance scheduling}
In this section, we summarize the proposed formulation of safe RL in Algorithm \ref{Algorithm1}. This algorithm shows how all units can find their maintenance scheduling and electricity market bidding.
\begin{algorithm}
	\caption{Safe RL algorithm for bidding and maintenance scheduling} 
	\label{Algorithm1}
	\begin{algorithmic}[1] 
		\STATE {\textbf{Input}}: $\gamma,\biddingmax,H_{i},D_{i},\powermin,\powermax,\rampup,\rampdown,\theta_{i},\theta'_{i},\eta_{i},\tau_{i},\theta^{\mu}_{i},\theta^{'\mu}_{i},\eta^{\mu}_{i},\tau^{\mu}_{i},M,N_{s}, \N$\\ 
		        \STATE Initialize the system safety state $x_{f,i}(1)=0$, $\I$
        \FOR{$episode = 1,\ldots,\mathcal{M}$}
        \STATE Initialize the system state $s(t)$ using \eqref{eq:state}
        \STATE $\mathcal{T}={\{1+(episode-1)T,\cdots,(episode)T}\}$
        \FOR{$t \in\mathcal{T}$}
        \STATE Obtain the action from the actor network $a_{i}(t)=\max(\min(\mu_{i}(s(t),\theta_{i}^{\mu}),\biddingmax),1)$,\quad $\N$
        \STATE Obtain the maintenance scheduling which satisfies the safety filter $u_{f,i}(t)$ using \eqref{eq:safety}
        \STATE $u_{i}(t)=u_{f,i}(t)$,\quad $\N$
        \STATE Obtain the market clearing price $\lambda(t)$ and units' power generation $g_{i}(t)$, $\N$, using \eqref{eq:ISO}
        \STATE Observe the next state $s(t+1)=(\lambda(t),d(t))$ 
        \STATE Calculate the reward function $r_{i}(t)$, $\N$, using \eqref{eq:reward}
        \STATE Store the transition $(s(t),s(t+1),a_{i}(t),r_{i}(t))$, $\N$
        \STATE Randomly sample $(s^{j}(t),s^{j}(t+1),a^{j}_{i}(t),r^{j}_{i}(t))$, $\N$, $\J$
        \STATE Obtain the target network $Q_{i,target}^{j}(t)$, $\N$, $\J$, using \eqref{eq:q_target}
        \STATE Update the critic networks' weights $\theta_{i}$, $\N$, using \eqref{eq:critic weight}
        \STATE Update the actor networks' weights $\theta_{i}^{\mu}$, $\N$, using \eqref{eq:actor network1}
        \STATE Update the target critic networks' weights $\theta'_{i}$, $\N$, using \eqref{eq:target critic network}
         \STATE Update the target actor networks' weights $\theta_{i}^{'\mu}$, $\N$, using \eqref{eq:actor target network}
        \ENDFOR
                \ENDFOR
	\end{algorithmic}
\end{algorithm}

\section{Case study evaluation results}
\label{sec:result}

In this section, we evaluate the performance of the proposed algorithm on IEEE 30-bus system with $6$ generation units \cite{bompard2006network}. We consider a linear marginal cost for each generation and neglect the ramp up and down of the generating units. The maximum bidding $\biddingmax$ is considered as being $2$ (Constraint $A_1$ in \ref{eq:agent model}). Moreover, to increase the safety of the units, we impose the constraint that each unit should perform maintenance for at least $1$ day during $100$ days (Constraint $A_{2}$ in \eqref{eq:agent model}). In addition, the repair duration time of a single maintenance action for all of generating units $D_{i}$ is considered to be one day. It means that once they decide to perform maintenance, it should be taken at least one day (Constraint $A_3$ in \eqref{eq:agent model}). We assume that the maximum number of generating units that can perform maintenance simultaneously $M$ is $2$ (Constraint $S_3$ in \eqref{eq:safety}). This limitation makes sure that the load of the system can be satisfied at all periods of time.  The other parameters of the generating units are shown in Table \ref{tab:generation}. The parameters of the network are obtained from \cite{bompard2006network}.

\begin{table}
  \caption{Generation units' parameters}
  \centering
 \begin{tabular}{c | c c c c} 
Generation unit&  $\marginalcost$  & $\powermax$ &   $\powermin$ &$c_{i}$ \\
& $[\frac{\$}{MWh}]$  & $[MW]$  & $[MW]$ & $[{\$}]$ \\
 \hline
 1 &  2& 80 &5 & 120 \\ 
 \hline
 2 & 1.75 & 80 &5 &135  \\
 \hline
 3 & 1& 50 &5 &142 \\
 \hline
 4 &  3.25& 55& 5 &125\\
 \hline
 5 & 3 &30 & 5 &175\\
  \hline
 6 & 3 &40 & 5 &165 \\
\end{tabular}
  \label{tab:generation}
\end{table}
In this case study to implement Algorithm \ref{Algorithm1}, we consider $100$ episodes in which each episode consists of $30$ days. We set two hidden layers for the actor-critic networks of each generating unit. In each layer, we use the Rectified Linear Unit (ReLU ) activation function. The initial value for the weights of neural networks is chosen randomly between 1 and 3. The number of samples $N_{s}=100$ is chosen for the DDPG algorithm. In the following, we evaluate the performance of the safe RL Algorithm \ref{tab:generation}.

\subsection{Generation unit reward}
\begin{figure}[!t]
\centerline{\includegraphics[width=\columnwidth]{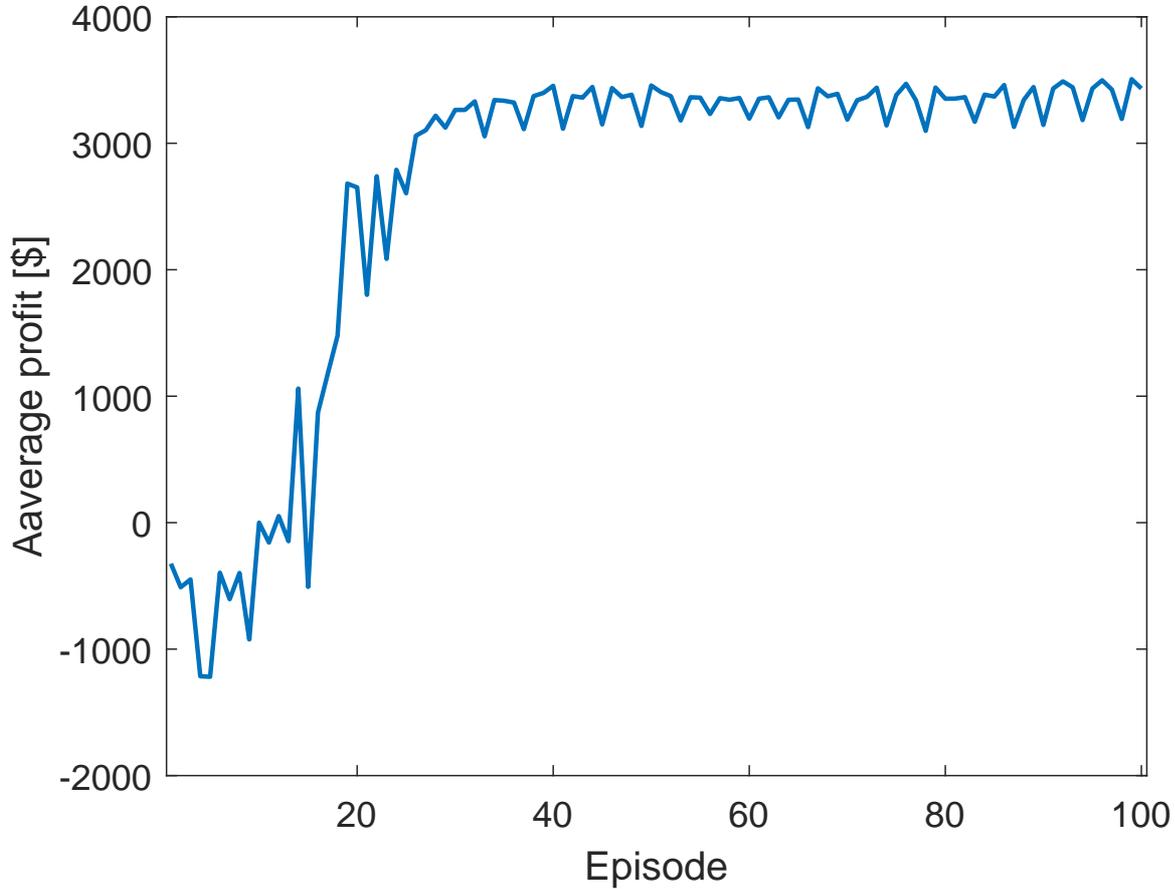}}
\caption{Sum of the average reward over each episode for all units using the learning algorithm.}
\label{fig:profit}
\end{figure}

We obtain the average reward over each episode for all units ($\bar{r}_{i}(episode)=\frac{\sum\limits_{{1+(episode-1)T}}^{(episode)T}{r_{i}(t)}}{T}$ $,\I$, $episode=1,\cdots,\mathcal{M}$). Then
We calculate the summation of this average reward over all units ($\SUMI{\bar{r}_{i}(episode)}$) and depict it for all episode ($episode=1,\cdots,\mathcal{M}$) in Figure \ref{fig:profit}. As we can see in this figure, at the beginning, the average reward is comparably low since the units are not aware of the market price and they are in the training stage. This figure shows that the reward is increasing during training which means that the units can increase their profit by changing their bidding and maintenance strategies. The algorithm converges after about $40$ episodes.

\begin{figure}[!t]
\centerline{\includegraphics[width=\columnwidth]{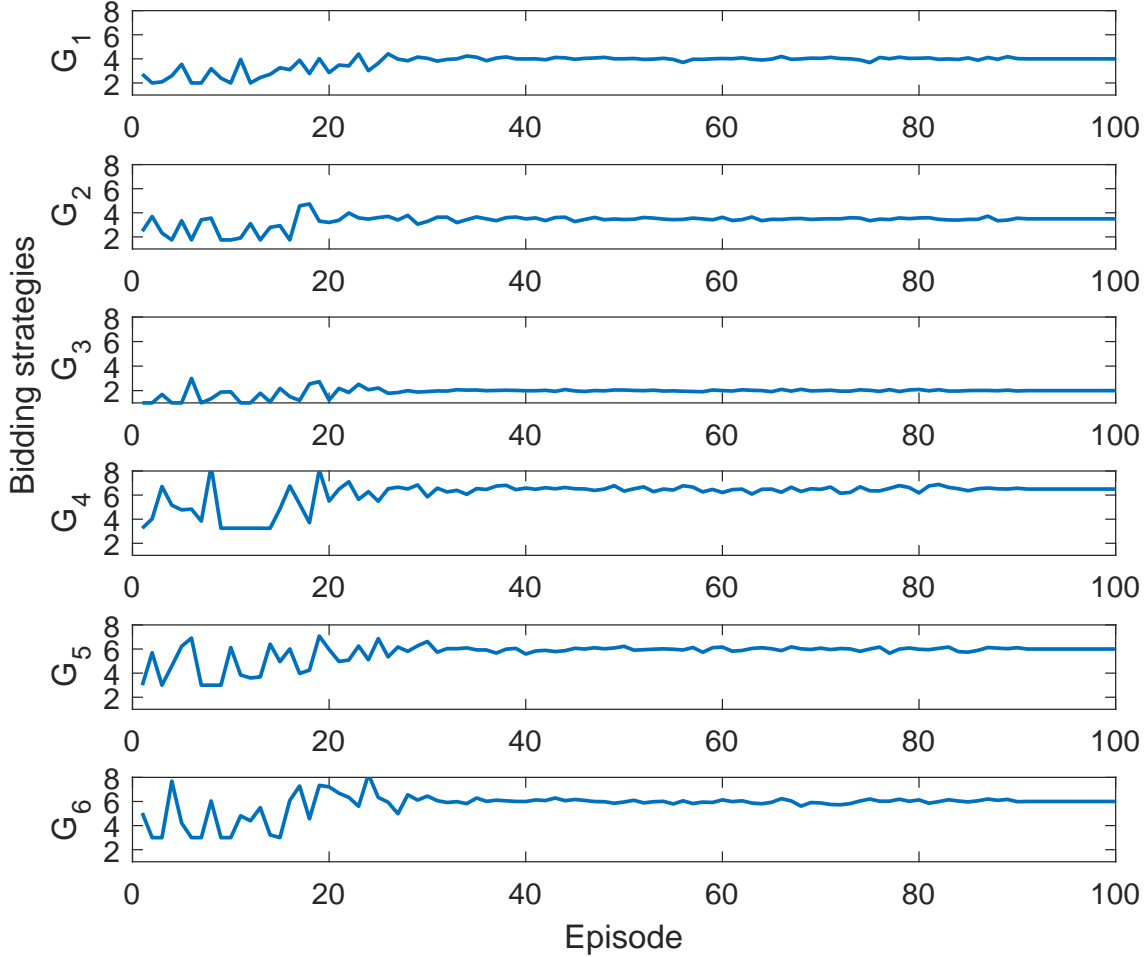}}
\caption{Average biding of all units during each episode using the learning algorithm.}
\label{fig:bidding}
\end{figure}

The bidding strategies of units $\biddingt$, $\I$, are shown in Figure \ref{fig:bidding}. This figure shows that the bidding of units converges after about $40$ episodes of the algorithm and that the units learn how to change their bidding to improve their profit. 

\subsection{Maintenance scheduling}
The maintenance scheduling of generation units between day $700$ until $900$ is shown in Figure \ref{fig:maintenance} (we only show a limited time interval to visualize the frequency of maintenance scheduling more clearly). This figure shows that at each time step, the maximum of two generating units can perform maintenance simultaneously. Moreover, each generation performs maintenance at least once during 100 days. This increases the system reliability and safety.

\begin{figure}[!t]
\centerline{\includegraphics[width=\columnwidth]{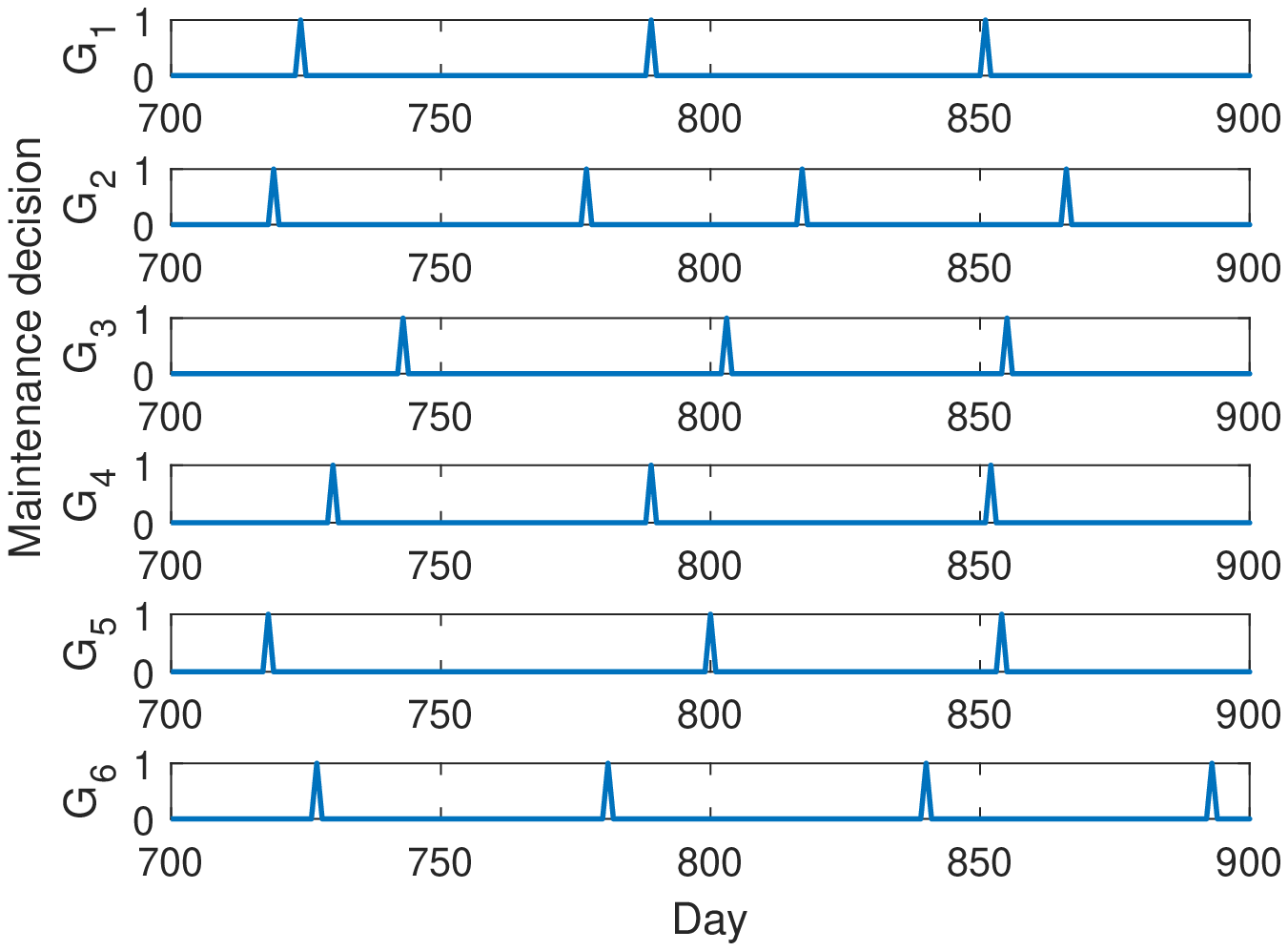}}
\caption{Maintenance scheduling of generation units during the learning algorithm between day $700$ until $900$.}
\label{fig:maintenance}
\end{figure}

The average maintenance cost per unit for each episode and during the learning algorithm is demonstrated in Figure \ref{fig:maintenance}. According to this figure, the maintenance cost is decreasing during the learning algorithm which means that units learn how to find the sub-optimal maintenance scheduling and do not perform maintenance frequently.

\begin{figure}[!t]
\centerline{\includegraphics[width=\columnwidth]{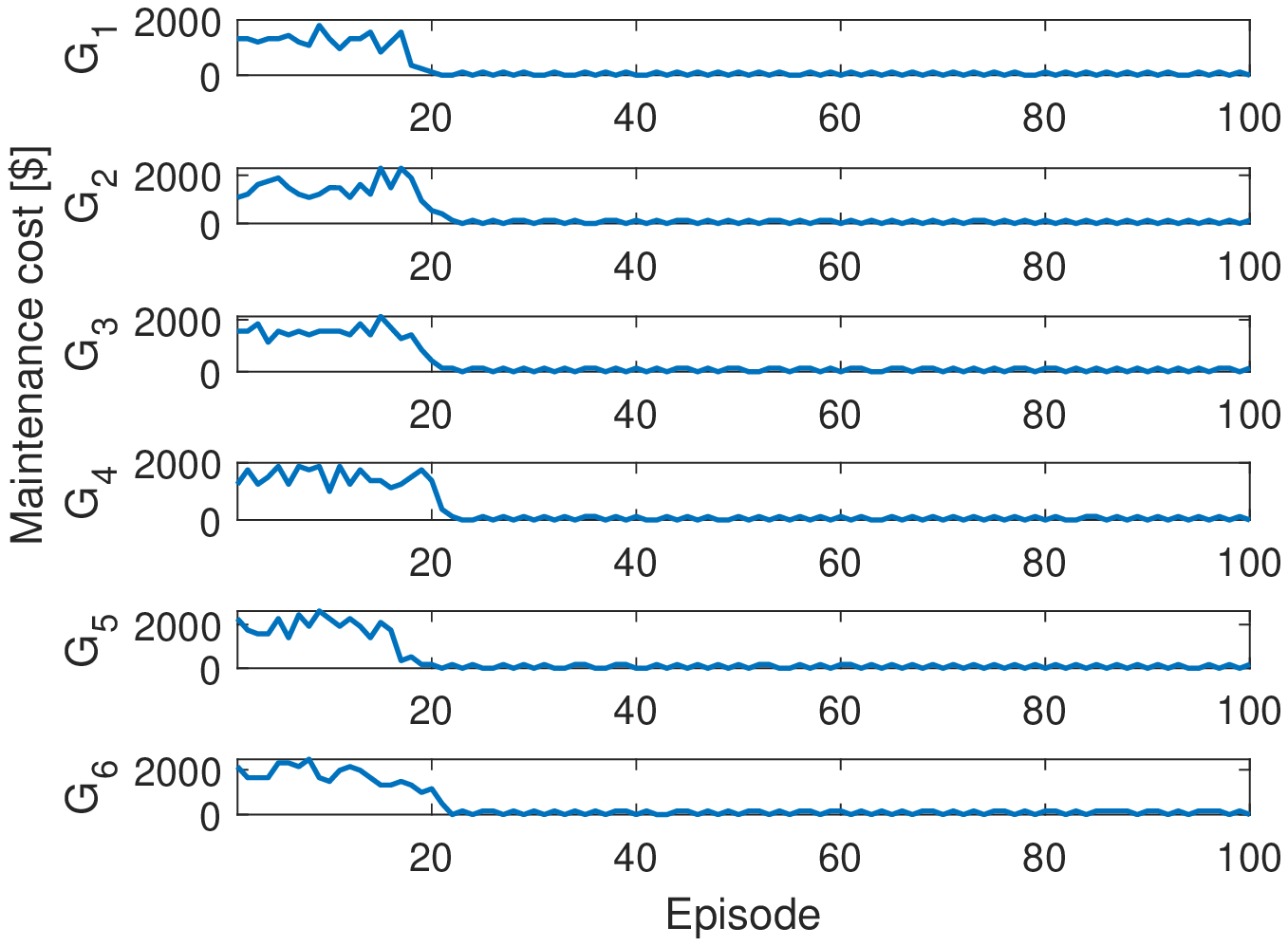}}
\caption{Average maintenance cost during each episode for all units using Algorithm \ref{Algorithm1}.}
\label{fig:maintenance}
\end{figure}


\subsection{Effect of the safety filter}
\begin{figure}[!t]
\centerline{\includegraphics[width=\columnwidth]{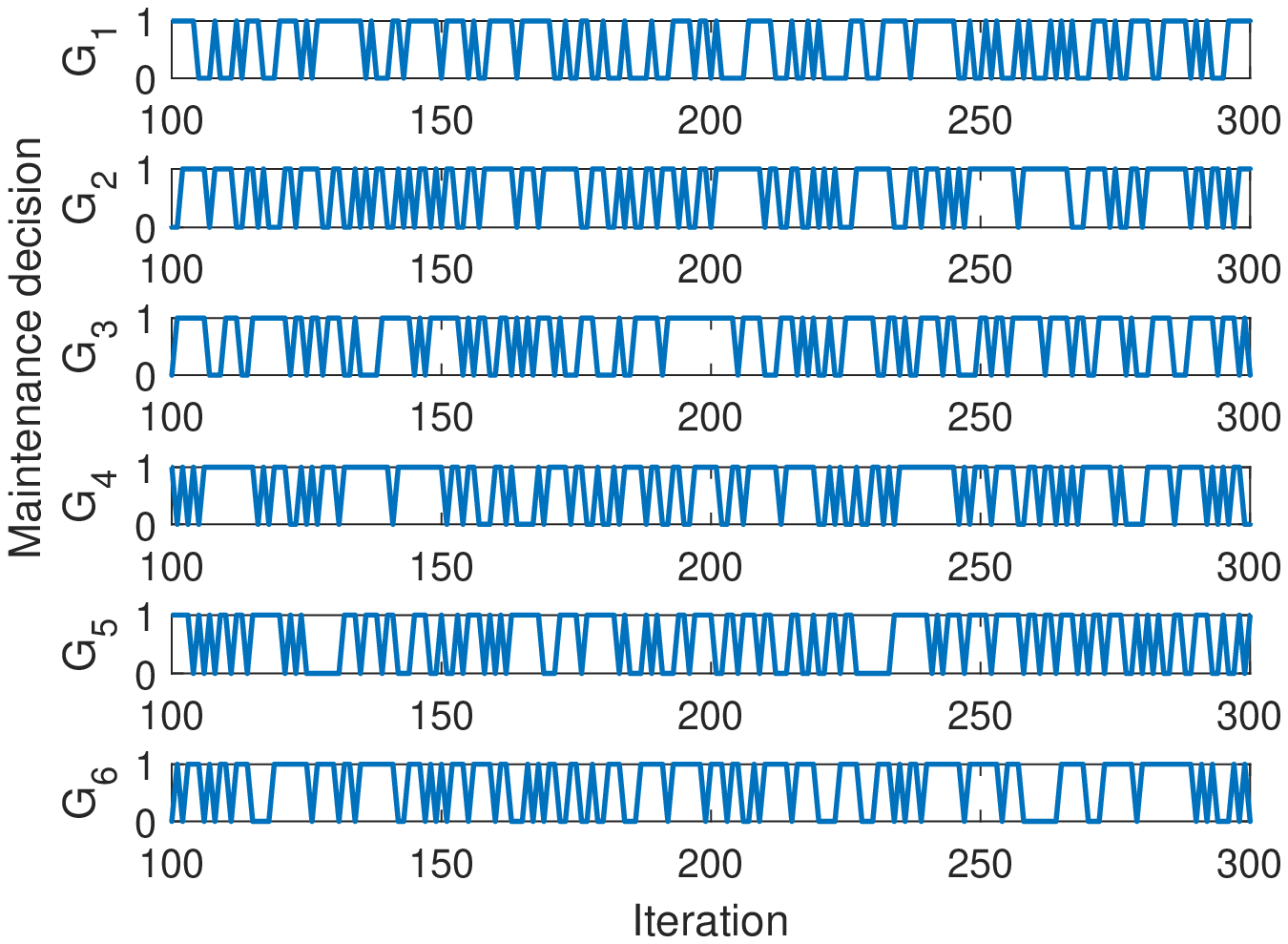}}
\caption{Maintenance scheduling of units without presence of safety filter during the day $100$ until $300$.}
\label{fig:safety load}
\end{figure}

As mentioned in Section \ref{sec:solution}, the RL algorithm obtains the units' strategies without considering the safety constraints. The predicted safety filter is then responsible to satisfy the safety constraints. In the case that we do not consider the safety filter, there is not any guarantee that the units perform maintenance before they fail. For-example, during the interval between iteration $700$ and $900$, the units do not perform maintenance which decreases the system safety. In addition, during training, when units are not aware of the other units' maintenance decision, did not learn their best strategies, and explore different strategies to learn the behaviour of system, using only RL without the safety filter might result in the case that most of the units may schedule their maintenance simultaneously. In this case, the demand of the system would not be able to be satisfied. Figure \ref{fig:safety load} shows the time interval where most of the units perform maintenance simultaneously and there is not any guarantee for satisfying the load of the system.

\subsection{Performance comparison of different RL algorithms}
\begin{figure}[!t]
\centerline{\includegraphics[width=\columnwidth]{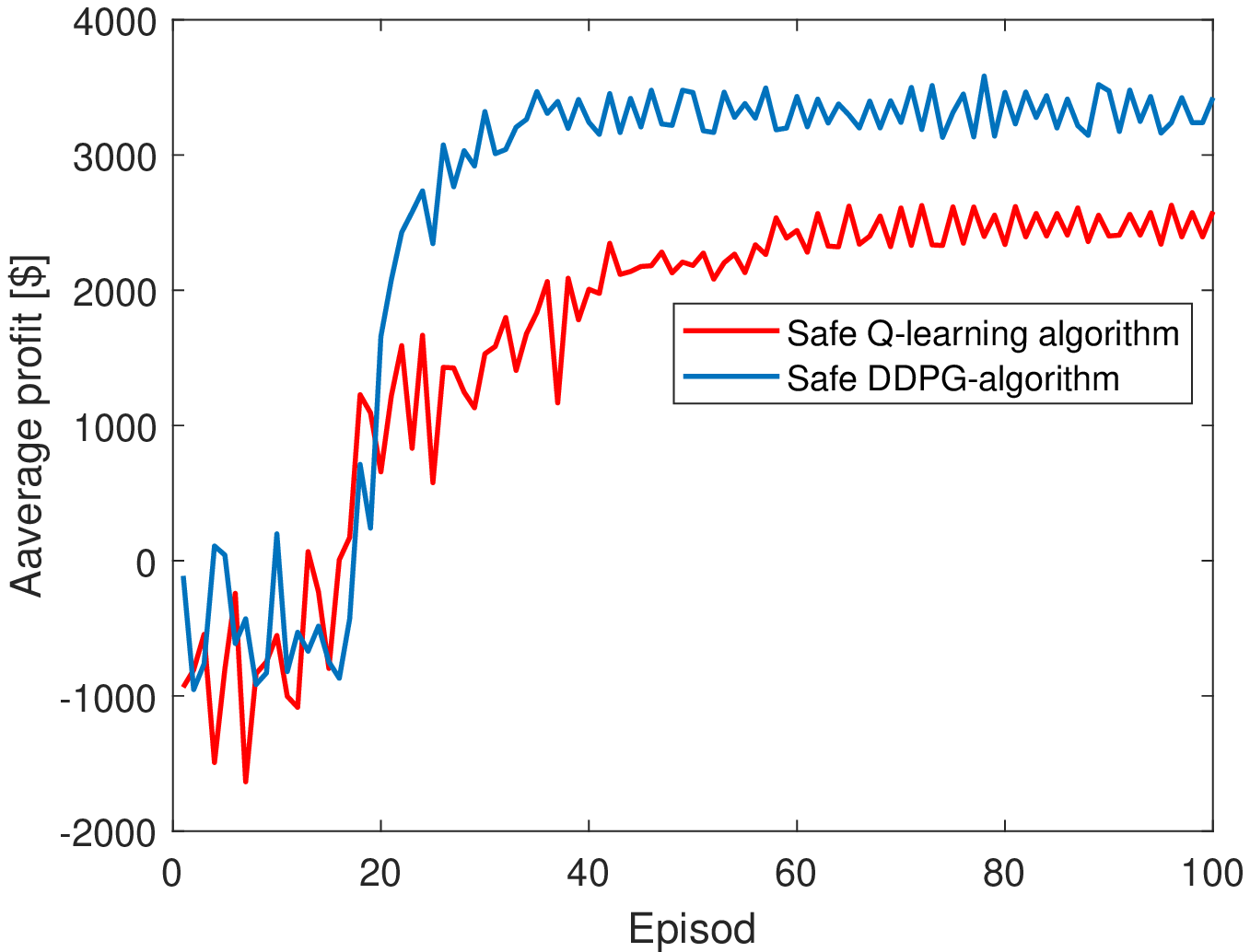}}
\caption{Comparison of the average profit over all units of safe Q-learning with safe DDPG algorithm.}
\label{fig:compare}
\end{figure}

In this section, we compare the results of Algorithm \ref{Algorithm1} with the Q-learning algorithm. In Q-learning, the RL algorithm applies a look-up table to update its policy. To obtain the look-up table, the states and actions of the system should be discretized. Hence, we consider discrete points for price and load as the state and bidding strategy as the action. Figure \ref{fig:compare} shows the average profit of the safe DDPG algorithm and the safe Q-learning algorithm.  As we can see, the DDPG algorithm converges in fewer episodes than the Q-learning. Moreover, the DDPG algorithm can achieve a higher profit than the Q-learning algorithm. These results can be explained with the fact that in this problem, the state and the actions are continuous. Hence, using a look-up table which approximates the continuous space with some discrete values decreases the performance and accuracy of the algorithm.

\section{Conclusions}
\label{sec:conclusion}
In this paper, we address the problem of joint bidding and maintenance schedule in the electricity market environment. We propose the safe deep RL algorithm to solve the problem. In the first step of this algorithm, each unit obtains its strategies using deep RL methodology without considering the safety constraints. Then in the second step, the safety filter modifies the units' decisions to ensure that all the system constraints can be satisfied. The proposed algorithm can handle the challenges of uncertainty and safety constraints. The results on the case study show that during the training, the profit of the units is increasing, while the generation units also perform preventive maintenance. In addition, the results demonstrate that the proposed methodology can achieve a higher profit compared to the Q-learning algorithm. 

As future work, one can consider some other sources of uncertainties in the electricity market such as renewable energies. In addition, the proposed algorithm can be applied to a larger system with more generation units by considering additional constraints such as contingency of the network.

\bibliography{bib_item}
\bibliographystyle{ieeetr}

\end{document}

%% file: style_P.tex
\makeatletter
\def\@settitle{\begin{center}%
		\baselineskip14\p@\relax
		\normalfont\LARGE\scshape\bfseries
		\@title
	\end{center}%
}
\makeatother

\makeatletter

\def\subsection{\@startsection{subsection}{2}%
	\z@{.5\linespacing\@plus.7\linespacing}{.5\linespacing}%
	{\normalfont\bfseries}}

\def\subsubsection{\@startsection{subsubsection}{3}%
	\z@{.5\linespacing\@plus.7\linespacing}{.5\linespacing}%
	{\normalfont\itshape}}

\usepackage[usenames, dvipsnames]{color}
\definecolor{darkblue}{rgb}{0.0, 0.0, 0.45}

\usepackage[colorlinks	= true,
			raiselinks	= true,
			linkcolor	= darkblue, 
			citecolor	= Mahogany,
			urlcolor	= ForestGreen,
			pdfauthor	= {},
			pdftitle	= {},
			pdfkeywords	= {},
			pdfsubject	= {},
			plainpages	= false]{hyperref}

\usepackage{dsfont,amssymb,amsmath,graphicx} 
\usepackage{amsfonts,dsfont,mathtools,mathrsfs,amsthm} 
\mathtoolsset{showonlyrefs=true}
\usepackage[amssymb, thickqspace]{SIunits}
\usepackage{fancyhdr,setspace}

\usepackage[numbers,compress]{natbib}
\usepackage{nicefrac}       
\usepackage{microtype}      
\usepackage{tabto}

\usepackage{enumerate,enumitem}
 \usepackage{caption}

\usepackage{bbm}

\allowdisplaybreaks
\date{\today}

%% file: command.tex
\theoremstyle{theorem}

\theoremstyle{remark}

\newcommand{\powermax}{g_{i}^{\max}}
\newcommand{\powermin}{g_{i}^{\min}}
\newcommand{\powerp}{g_{i}(t-1)}

\newcommand{\marginalcost}{\lambda_{i,m}}

\newcommand{\price}{\lambda(t)}
\newcommand{\maintenance}{u_{i}}
\newcommand{\bidding}{k_{i}}
\newcommand{\biddingt}{k_{i}(t)}

\newcommand{\biddingmax}{k_{i}^{\max}}

\newcommand{\maintenancet}{u_{i}(t)}
\newcommand{\maintenancetp}{u_{i}(t-1)}

\newcommand{\maintenanced}{u_{i}(t+D_{i}-1)}

\newcommand{\SUM}{\sum\limits_{t\in\mathcal{T}}}
\newcommand{\TA}{\forall{t\in\mathcal{T}}}

\newcommand{\SUMI}{\sum\limits_{i\in\mathcal{N}}}

\newcommand{\rampup}{R_{i}^{U}}
\newcommand{\rampdown}{R_{i}^{D}}

\newcommand{\state}{x_{f,i}(t')}
\newcommand{\statenext}{x_{f,i}(t'+1)}
\newcommand{\statefinal}{x_{f,i}(T)}
\newcommand{\actionfilter}{u_{f,i}}
\newcommand{\actionfiltert}{u_{f,i}(t)}
\newcommand{\J}{\forall j\in\mathcal{N}_{s}}
\newcommand{\I}{\forall i\in\mathcal{N}}
\newcommand{\TS}{t'={\{t,\cdots,T}\}}
\newcommand{\N}{{\forall i\in\mathcal{N}}}

%% file: main.bbl
\begin{thebibliography}{10}

\bibitem{wang2017strategic}
C.~Wang, W.~Wei, J.~Wang, F.~Liu, and S.~Mei, ``Strategic offering and
  equilibrium in coupled gas and electricity markets,'' {\em IEEE Transactions
  on Power Systems}, vol.~33, no.~1, pp.~290--306, 2017.

\bibitem{du2021approximating}
Y.~Du, F.~Li, H.~Zandi, and Y.~Xue, ``Approximating nash equilibrium in
  day-ahead electricity market bidding with multi-agent deep reinforcement
  learning,'' {\em Journal of Modern Power Systems and Clean Energy}, no.~99,
  pp.~1--11, 2021.

\bibitem{kohansal2020strategic}
M.~Kohansal, A.~Sadeghi-Mobarakeh, S.~D. Manshadi, and H.~Mohsenian-Rad,
  ``Strategic convergence bidding in nodal electricity markets: optimal bid
  selection and market implications,'' {\em IEEE Transactions on Power
  Systems}, vol.~36, no.~2, pp.~891--901, 2020.

\bibitem{zhang2010competitive}
G.~Zhang, G.~Zhang, Y.~Gao, and J.~Lu, ``Competitive strategic bidding
  optimization in electricity markets using bilevel programming and swarm
  technique,'' {\em IEEE Transactions on Industrial Electronics}, vol.~58,
  no.~6, pp.~2138--2146, 2010.

\bibitem{yildirim2016sensor}
M.~Yildirim, X.~A. Sun, and N.~Z. Gebraeel, ``Sensor-driven condition-based
  generator maintenance scheduling—part i: Maintenance problem,'' {\em IEEE
  Transactions on Power Systems}, vol.~31, no.~6, pp.~4253--4262, 2016.

\bibitem{chen1991optimal}
L.~Chen and J.~Toyoda, ``Optimal generating unit maintenance scheduling for
  multi-area system with network constraints,'' {\em IEEE Transactions on Power
  Systems}, vol.~6, no.~3, pp.~1168--1174, 1991.

\bibitem{volkanovski2008genetic}
A.~Volkanovski, B.~Mavko, T.~Bo{\v{s}}evski, A.~{\v{C}}au{\v{s}}evski, and
  M.~{\v{C}}epin, ``Genetic algorithm optimisation of the maintenance
  scheduling of generating units in a power system,'' {\em Reliability
  Engineering \& System Safety}, vol.~93, no.~6, pp.~779--789, 2008.

\bibitem{song2002nash}
H.~Song, C.-C. Liu, and J.~Lawarr{\'e}e, ``Nash equilibrium bidding strategies
  in a bilateral electricity market,'' {\em IEEE transactions on Power
  Systems}, vol.~17, no.~1, pp.~73--79, 2002.

\bibitem{ye2019incorporating}
Y.~Ye, D.~Papadaskalopoulos, J.~Kazempour, and G.~Strbac, ``Incorporating
  non-convex operating characteristics into bi-level optimization electricity
  market models,'' {\em IEEE Transactions on Power Systems}, vol.~35, no.~1,
  pp.~163--176, 2019.

\bibitem{dai2016finding}
T.~Dai and W.~Qiao, ``Finding equilibria in the pool-based electricity market
  with strategic wind power producers and network constraints,'' {\em IEEE
  Transactions on Power Systems}, vol.~32, no.~1, pp.~389--399, 2016.

\bibitem{zhang2010restructured}
X.-P. Zhang, {\em Restructured electric power systems: analysis of electricity
  markets with equilibrium models}, vol.~71.
\newblock John Wiley \& Sons, 2010.

\bibitem{pozo2011finding}
D.~Pozo and J.~Contreras, ``Finding multiple nash equilibria in pool-based
  markets: A stochastic epec approach,'' {\em IEEE Transactions on Power
  Systems}, vol.~26, no.~3, pp.~1744--1752, 2011.

\bibitem{marwali1999long}
M.~Marwali and S.~Shahidehpour, ``Long-term transmission and generation
  maintenance scheduling with network, fuel and emission constraints,'' {\em
  IEEE Transactions on power systems}, vol.~14, no.~3, pp.~1160--1165, 1999.

\bibitem{pandzic2011yearly}
H.~Pandzic, A.~J. Conejo, I.~Kuzle, and E.~Caro, ``Yearly maintenance
  scheduling of transmission lines within a market environment,'' {\em IEEE
  Transactions on Power Systems}, vol.~27, no.~1, pp.~407--415, 2011.

\bibitem{feng2009competitive}
C.~Feng and X.~Wang, ``A competitive mechanism of unit maintenance scheduling
  in a deregulated environment,'' {\em IEEE transactions on power systems},
  vol.~25, no.~1, pp.~351--359, 2009.

\bibitem{min2013game}
C.~Min, M.~Kim, J.~Park, and Y.~Yoon, ``Game-theory-based generation
  maintenance scheduling in electricity markets,'' {\em Energy}, vol.~55,
  pp.~310--318, 2013.

\bibitem{rokhforoz2021multi}
P.~Rokhforoz, B.~Gjorgiev, G.~Sansavini, and O.~Fink, ``Multi-agent maintenance
  scheduling based on the coordination between central operator and
  decentralized producers in an electricity market,'' {\em Reliability
  Engineering \& System Safety}, vol.~210, p.~107495, 2021.

\bibitem{conejo2005generation}
A.~J. Conejo, R.~Garc{\'\i}a-Bertrand, and M.~D{\'\i}az-Salazar, ``Generation
  maintenance scheduling in restructured power systems,'' {\em IEEE
  Transactions on Power Systems}, vol.~20, no.~2, pp.~984--992, 2005.

\bibitem{sutton2018reinforcement}
R.~S. Sutton and A.~G. Barto, {\em Reinforcement learning: An introduction}.
\newblock MIT press, 2018.

\bibitem{watkins1992q}
C.~J. Watkins and P.~Dayan, ``Q-learning,'' {\em Machine learning}, vol.~8,
  no.~3-4, pp.~279--292, 1992.

\bibitem{millan2002continuous}
J.~D.~R. Mill{\'a}n, D.~Posenato, and E.~Dedieu, ``Continuous-action
  q-learning,'' {\em Machine Learning}, vol.~49, no.~2, pp.~247--265, 2002.

\bibitem{arulkumaran2017deep}
K.~Arulkumaran, M.~P. Deisenroth, M.~Brundage, and A.~A. Bharath, ``Deep
  reinforcement learning: A brief survey,'' {\em IEEE Signal Processing
  Magazine}, vol.~34, no.~6, pp.~26--38, 2017.

\bibitem{xu2019deep}
H.~Xu, H.~Sun, D.~Nikovski, S.~Kitamura, K.~Mori, and H.~Hashimoto, ``Deep
  reinforcement learning for joint bidding and pricing of load serving
  entity,'' {\em IEEE Transactions on Smart Grid}, vol.~10, no.~6,
  pp.~6366--6375, 2019.

\bibitem{liang2020agent}
Y.~Liang, C.~Guo, Z.~Ding, and H.~Hua, ``Agent-based modeling in electricity
  market using deep deterministic policy gradient algorithm,'' {\em IEEE
  Transactions on Power Systems}, vol.~35, no.~6, pp.~4180--4192, 2020.

\bibitem{ye2019deep}
Y.~Ye, D.~Qiu, M.~Sun, D.~Papadaskalopoulos, and G.~Strbac, ``Deep
  reinforcement learning for strategic bidding in electricity markets,'' {\em
  IEEE Transactions on Smart Grid}, vol.~11, no.~2, pp.~1343--1355, 2019.

\bibitem{wabersich2021predictive}
K.~P. Wabersich and M.~N. Zeilinger, ``A predictive safety filter for
  learning-based control of constrained nonlinear dynamical systems,'' {\em
  Automatica}, vol.~129, p.~109597, 2021.

\bibitem{menache2005basis}
I.~Menache, S.~Mannor, and N.~Shimkin, ``Basis function adaptation in temporal
  difference reinforcement learning,'' {\em Annals of Operations Research},
  vol.~134, no.~1, pp.~215--238, 2005.

\bibitem{Fortuny_1981}
A.~Fortuny and B.~McCarl, ``A representation and economic interpretation of a
  two-level programming problem,'' {\em Journal of the operational Research
  Society}, vol.~32, no.~9, pp.~783--792, 1981.

\bibitem{bompard2006network}
E.~Bompard, W.~Lu, and R.~Napoli, ``Network constraint impacts on the
  competitive electricity markets under supply-side strategic bidding,'' {\em
  IEEE Transactions on Power Systems}, vol.~21, no.~1, pp.~160--170, 2006.

\end{thebibliography}
